# Thermodynamic evidence of fractionalized excitations in α-RuCl$_3$

S. Widmann,[1] V. Tsurkan,[1,2] D. A. Prishchenko,[3] V. G. Mazurenko,[3] A. A. Tsirlin,[4] and A. Loidl[1*]

[1]*Experimental Physics V, Center for Electronic Correlations and Magnetism, University of Augsburg, 86159 Augsburg, Germany*
[2]*Institute of Applied Physics, MD 2028 Chisinau, Republic of Moldova*
[3]*Ural Federal University, Mira Street 19, 620002 Ekaterinburg, Russia*
[4]*Experimental Physics VI, Center for Electronic Correlations and Magnetism, University of Augsburg, 86159 Augsburg, Germany*

Fractionalized excitations are of considerable interest in recent condensed-matter physics. Fractionalization of the spin degrees of freedom into localized and itinerant Majorana fermions are predicted for the Kitaev spin liquid, an exactly solvable model with bond-dependent interactions on a two-dimensional honeycomb lattice. As function of temperature, theory predicts a characteristic two-peak structure of the heat capacity as fingerprint of these excitations. Here we report on detailed heat-capacity experiments as function of temperature and magnetic field in high-quality single crystals of α-RuCl$_3$ and undertook considerable efforts to determine the exact phonon background. We measured single-crystalline RhCl$_3$ as non-magnetic reference and performed ab-initio calculations of the phonon density of states for both compounds. These ab-initio calculations document that the intrinsic phonon contribution to the heat capacity cannot be obtained by a simple rescaling of the nonmagnetic reference using differences in the atomic masses. Sizable renormalization is required even for non-magnetic RhCl$_3$ with its minute difference from the title compound. In α-RuCl$_3$ in zero magnetic field, excess heat capacity exists at temperatures well above the onset of magnetic order. In external magnetic fields far beyond quantum criticality, when long-range magnetic order is fully suppressed, the excess heat capacity exhibits the characteristic two-peak structure. In zero field, the lower peak just appears at temperatures around the onset of magnetic order and seems to be connected with canonical spin degrees of freedom. At higher fields, beyond the critical field, this peak is shifted to 10 K. The high-temperature peak located around 50 K is hardly influenced by external magnetic fields, carries the predicted amount of entropy, $R/2 \ln 2$, and may resemble remnants of Kitaev physics.

## I. INTRODUCTION

α-RuCl$_3$ is a prime candidate to exhibit a Kitaev type spin-liquid ground state. The Kitaev model [1] represents a spin $S = ½$ system on a two-dimensional (2D) honeycomb lattice with bond-dependent interactions. The ground state of this model is exactly solvable and predicts the existence of gapped and gapless excitations, depending on the asymmetry of the exchange interactions. The exact solution is provided by fractionalization of quantum spins $S = ½$ into two types of Majorana fermions: $Z_2$ fluxes and itinerant fermions. The former being localized on each hexagon of the honeycomb lattice, the latter forming propagating fermionic quasiparticles. Theory predicts clear hallmarks of these fractionalized spin degrees of freedom in the temperature-dependent heat capacity on well-separated temperature scales. In the ground state, the $Z_2$ fluxes are frozen in the topologically ordered quantum-spin liquid (QSL) and fluctuating fluxes excite localized Majorana fermions. On further increasing temperatures, itinerant quasiparticles become activated. These two classes of excitations release their entropy successively in two steps at vastly different temperatures. Details of the Kitaev physics and appropriate models and materials are discussed in two recent review articles [2,3].

In the concept of QSLs [4] it is implicitly supposed that the paramagnetic phase is continuously connected with the QSL state and, to our knowledge, a thermodynamic phase transition has never been reported experimentally. Hence, it seems of prime importance to search for these thermody-namic anomalies when entering the Kitaev spin liquid (KSL). The first to estimate the characteristic thermodynamic temperature dependence of fractionalized excitations were Nasu et al. [5,6] and they indeed found characteristic anomalies separating the paramagnet from the QSL. Using Quantum Monte Carlo methods, they calculated the temperature dependence of the specific heat $C$ and the corresponding entropy $S$. Using a model involving pure Kitaev exchange $K$ only, they found a characteristic two-peak spectrum of the heat capacity, separated in temperature by at least one order of magnitude. Each of the peaks is characterized by an entropy release of $R/2 \ln 2$, with $R$ being the gas constant. In addition, the release of itinerant Majorana fermions is correlated with a linear increase of the heat capacity. A cluster dynamical mean-field study for the Kitaev model of the temperature dependent heat capacity with a similar result has been performed by Yoshitake et al. [7]. A two-peak structure in $C(T)$ has even been predicted in systems with magnetic order in proximity to a KSL phase [8].

Utilizing a model with ferromagnetic Kitaev and antiferromagnetic symmetric off-diagonal exchange $\Gamma$, Catuneanu et al. [9] and Samarakoon et al. [10] calculated susceptibilities and thermodynamic quantities of the Kitaev spin-liquid candidate α-RuCl$_3$. The main motivation was to explain the scattering continua as observed by neutron [11,12,13,14] and Raman scattering experiments [15,16,17] as well as by THz spectroscopy [18,19], but these authors also calculated the temperature dependence of heat capacity and found the characteristic two-peak structure and the concomitant two-step evolution of entropy. With the characteristic exchange parameters of $K$ and $\Gamma$ of the title compound, according to these authors, the two peaks in the heat capacity are located close to 0.03 $\Gamma$ and 0.4 $\Gamma$. Assuming an off-diagonal exchange of ~ 5 - 10 meV [20], two peaks in $C(T)$ will appear close to temperatures of ~ 2 K and ~ 30 K, respectively.

The temperature dependence of the heat capacity for α-RuCl$_3$ also was calculated by Suzuki and Suga [21] utilizing an exact numerical diagonalization method. Their results again reveal the characteristic two-peak structure with peaks close 2 and 20 K. Compared to experimental observations these temperatures seem to be significantly too low and the authors proposed an effective model involving strong ferromagnetic Kitaev coupling to quantitatively reproduce experimental results [21]. Pidatella et al. [22] have solved the Kitaev model in the presence of anisotropy. These authors found that the temperature scale for fermionic entropy release corresponds to temperatures comparable to the Kitaev exchange and does not depend on anisotropy, the temperature scale for the flux is strongly anisotropy dependent. It seems evident that this characteristic thermodynamic evidence of Majorana fermions should easily be tackled experimentally, if and only if the phonon properties of the model compounds are well known. It has recently been pointed out that even in systems with long-range or short-range magnetic order being close to a KSL state, magnon-like excitations being characteristic for conventional magnetic order give way to long-lived fractionalized quasi-particles at higher temperatures and energies [23].

At this point it has to be mentioned that the interpretation of experimental results in the title compound in terms of fractional excitations of the spin-liquid ground state remain controversial. It was proposed that the observed continua may represent incoherent excitations originating from strong magnetic anharmonicity [24,2] and a model using realistic exchange interactions was able to reproduce the evolution of the dynamical response at finite temperatures and magnetic fields [25].

Since the identification of α-RuCl$_3$ as KSL candidate, heat-capacity experiments have been routinely performed, in most cases however, in limited temperature regimes and not with a specific focus on spin fractionalization. Here we discuss these publications, mainly to show the problems related to a correct evaluation of the magnetic heat capacity by subtracting a plausible and realistic phononic background. Only in the latest published experiments, the involved scientists were aware of the predicted two-peak structure and tried to identify this hallmark of a KSL in their experiments.

First heat-capacity experiments were performed by Sears et al. [26], who observed anomalies due to subsequent magnetic phase transitions, with peaks located close to 7 and 10 K. These multiple magnetic phase transitions were related to inhomo-geneities in the sample, most probably due to the existence of a large number of stacking faults. The entropy of the peaks was calculated to be of order 0.8 J/(mol K), but of course, as stated by the authors, strongly will depend on the phonon subtraction. In this work, a polynomial fit in a limited temperature regime was used to determine the phonon background. Majumder et al. [27] measured the heat capacity of single-crystalline α-RuCl$_3$ as function of temperature and magnetic field $H$. These authors found two phase transitions close to 8 and 14 K and reported the suppression of both magnetic phase transitions for in-plane external magnetic fields > 9 T. Due to a missing proper phonon reference, the authors refrained from a detailed analysis of the low-temperature specific heat.

Cao et al. [28] provided a further detailed analysis of the low-temperature specific heat revealing a heat-capacity anomaly at the onset of antiferromagnetic (AFM) order close to 7 K. The phonon background was approximated utilizing a 2D Debye expression in a limited temperature range up to 20 K. $C(T)$ in single crystals up to 16 K has also been measured by Park et al. [29] and the specific heat anomaly at $T_N$ = 6.5 K was analyzed using a 2D Ising model. The magnetic contribution was estimated from the heat capacity of single crystalline ScCl$_3$ scaled to the appropriate mass of α-RuCl$_3$. $C(T,H)$ was investigated in detail by Kubota et al. [30], however on samples revealing a number of subsequent magnetic phase transitions. The magnetic specific heat was evaluated using nonmagnetic ScCl$_3$ as reference. In addition to the anomalies at the magnetic phase transitions, these authors found a broad maximum in $C/T$ around 85 K, which was interpreted to be caused by the short-range spin correlations. The magnetic entropy accumulated up to 140 K was of order $2R/3$ ln2. The small magnetic entropy was understood as being consistent with the low-spin state of Ru$^{3+}$ with effective spin $S$ = ½.

The heat capacity of α-RuCl$_3$ was reported by Hirobe et al. [31] in a work mainly dedicated to the analysis of thermal conductivity. The authors identified a peak in the temperature-dependent magnetic heat capacity close to 100 K and were the first to refer to the work of Nasu et al. [5,6] indicating that this peak could be related to the release in entropy when exciting itinerant Majorana fermions. So far, all these publications did not spend too much attention to the occurrence of the structural phase transition with significant hysteresis close to 150 K. The room-temperature symmetry of α-RuCl$_3$ is monoclinic with space group $C2/m$ [32,33,34] and undergoes a structural phase transition [17,18,29,30,35,36,37] probably into a low-temperature rhombohedral structure. The low-temperature crystallographic structure is not finally settled and other ground-state symmetries have been reported [38,39]. This uncertainty results from the multi-domain character of the low-temperature structure and from the fact that this layered van der Waals (vdW) system is prone to stacking faults. It seems unclear, how this structural phase transition influences heat capacity and the release of entropy specifically of the high-temperature anomaly of the KSL.

The first attempt to verify the theoretical predictions about the step-like release of entropy in α-RuCl$_3$ was undertaken by Do et al. [13]. These authors presented results of the temperature dependent heat capacity from 2 to 200 K. For a proper description of the phonon properties, the heat capacity of isostructural non-magnetic ScCl$_3$ was measured and scaled by the appropriate mass ratio. The residual magnetic heat capacity of the title compound showed the typical two-step release of entropy, each of order $R/2$ ln2, with maxima in the heat capacity close to 20 and 100 K. In addition, these authors identified the predicted linear increase of itinerant Majorana fermions for temperature from 50 to 100 K. The structural phase transition close to 170 K seemed to be of minor influence in the temperature range investigated. A critical remark concerning these measurements compared to theoretical predictions is the relatively high temperature of both peaks in the heat capacity, specifically that due to the localized fluxes, which theoretically was predicted at significantly lower temperatures and that the low-temperature peak is close to the onset of magnetic order.

There further exits a number of published low-temperature heat-capacity results, mainly dealing with the suppression of long-range magnetic order and the identification of a possible spin gap as function of external field. It is well documented in literature that in-plane external magnetic fields suppress AFM order close to 7 T, resulting in a quantum critical point [40,41,42,43]. All these references also treat and analyze the low-temperature specific heat. Ref. [40,43] use RhCl$_3$ as nonmagnetic reference. In Ref. [41] the heat capacity of nonmagnetic and isostructural α-IrCl$_3$ was taken to treat the phonon properties of the title compound. The focus of these experiments was on the scaling of the low-temperature specific heat and on the determination of a spin-excitation gap. However, one clearly can state that despite detailed investigations of the low-temperature heat capacity, in these above cited manuscripts there is no clear evidence of the low-temperature entropy release due to localized Majorana fluxes. Below 10 K and at low magnetic fields, the heat capacity is dominated by the anomaly due to the magnetic phase transition. In fields higher than the critical field, by some residues of these transition.

In the most detailed low-temperature investigation of the heat capacity by Wolter et al. [43], beyond quantum criticality a cusp-like anomaly appears ~ 10 K. However, the entropy release at this temperature is well below 1 J/(mol K) much smaller than $R/2 \ln 2$ expected in the Kitaev model. Very recently, the observation of a half-integer thermal quantum Hall effect was interpreted as hallmark of fractionalization of quantum spins [44]. Also in this work, the sub-Kelvin heat capacity was investigated at magnetic fields around the quantum-critical point. These results have not been corrected for possible phonon contributions.

In this work, we present detailed measurements of the specific heat as function of temperature and external magnetic field in high-quality single crystals of α-RuCl$_3$ and analyze our results using accurately determined phonon properties, which allowed to evidence excess magnetic contributions with high precision. Our results provide experimental evidence for the existence of excess heat capacity indicating a clear double-peak structure at all magnetic fields investigated. While we tend to interpret the low-temperature anomaly as being derived from conventional spin degrees of freedom, the high-temperature anomaly could indicate Majorana fermions of the KSL.

## II. EXPERIMENTAL DETAILS

High-quality α-RuCl$_3$ and isostructural nonmagnetic RhCl$_3$ single crystals used in this work were grown by vacuum sublimation. At room temperature, RhCl$_3$ crystallizes in the monoclinic $C2/m$ structure [45]. Details of sample growth and characterization of α-RuCl$_3$ are described in detail in Ref. [35,37]. The difference of the present sample batches as compared to previously investigated samples is the low magnetic ordering temperature of 6.4 K and the relatively narrow hysteresis of the structural phase transition, which extends from 131 to 163 K (see later). We interpret the present results as signaling high-quality single crystals, which undergo a well-defined monoclinic to rhombohedral structural phase transition and are characterized a by well-defined stacking sequence, with ABC type stacking in the high-temperature monoclinic structure and with AB stacking in the low-temperature rhombohedral phase, as has been observed in a number of isostructural layered chromium tri-halides [46,47,48,49].

Standard heat capacity was investigated in a Quantum Design PPMS for temperatures $1.8 < T < 250$ K and in magnetic fields up to 9 T. To identify the nature of the structural and magnetic phase transitions, canonical specific-heat experiments were supplemented by a large-pulse method, applying heat pulses leading up to a 20% increase of the sample temperature. Thereby the temperature response of the sample was analyzed separately on heating and cooling. In a second-order phase transition, both responses should be equal, while significant differences are expected at first-order phase transitions. The thermal expansion experiments were conducted with a Quantum Design PPMS equipped with a compact and miniaturized high-resolution capacitance dilatometer.

Phonon spectra and lattice contributions to the specific heat of α-RuCl$_3$ and RhCl$_3$ were calculated using the frozen-phonon method implemented in the PHONOPY code [50]. For details, see Appendix A.

## III. RESULTS AND DISCUSSION

To characterize the single crystals used in the present measurements, Figure 1(a) documents the temperature dependence of the heat capacity of α-RuCl$_3$ for temperatures from 2 K up to 250 K and for zero external magnetic fields (green open circles) as well as for fields of 9 T (orange open diamonds). At 0 T we find the characteristic λ-like heat-capacity anomaly signaling the transition into the AFM ground state at 6.4 K. In external in-plane fields, magnetic order becomes suppressed and at 9 T a remaining hump can be seen, slightly shifted to higher temperatures. A narrow peak at 163 K signals latent heat at a first-order structural phase transition, which clearly becomes visible only in large-pulse experiments and only on heating. As documented earlier [37], the structural phase transition is not influenced by magnetic fields up to 9 T.

Figure 1(b) shows the temperature dependence of the thermal expansion as measured along the crystallographic $c$ direction. A narrow well-developed hysteresis evolves between heating and cooling cycles, which spans a temperature range of 30 K only, compared to a very wide hysteresis reported in many experiments so far. This fact signals a well-developed structural phase transition where the stacking sequence changes in a relatively narrow temperature window. The inset documents a step-like increase of the thermal expansion at the onset of magnetic order. The length change along $c$ is significant and amounts ~ 100 ppm, signaling strong magneto-strictive effects, respectively strong spin-phonon coupling.

As documented in Fig. 1(a), the temperature dependence of the heat capacity has been determined with high precision over a wide temperature range. The important question that remains to be solved is, how the magnetic excess heat capacity can be determined with high-enough precision, to provide experimental evidence of fractionalized excitations. To solve this problem the appropriate phononic background has to be determined. But any phenomenological Debye-Einstein fit of the measured heat capacity of α-RuCl$_3$ will be hampered by the magnetic anomaly at low temperatures, by possible contributions of itinerant Majorana fermions to the heat capacity up to ~ 100 K and finally also by the structural phase transition at higher temperatures.

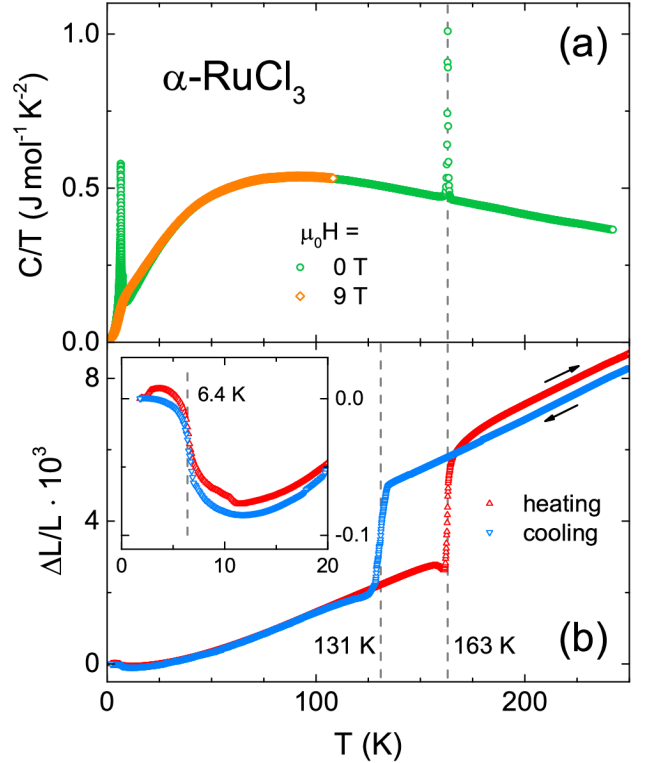

FIG. 1. (a) Temperature dependence of the heat capacity of α-RuCl$_3$ plotted as $C/T$, as observed in zero magnetic field (open green circles) and in-plane magnetic fields of 9 T (open orange diamonds). (b) Temperature dependence of the thermal expansion on heating (open red triangles up) and cooling (open blue triangles down). The measurements were performed along the crystallographic $c$ direction and were normalized at 1.8 K. A pronounced and well-defined hysteresis between heating and cooling cycle evolves between 131 and 163 K. The AFM phase transition is indicated by a significant step-like increase of the thermal expansion (see inset).

To get an estimate of the phonon background, we measured RhCl$_3$, which crystallizes in the same monoclinic room-temperature structure [45] and with rhodium as direct neighbor of Ru in the periodic table of elements with marginal differences of the atomic masses. Hence, while the mass relation are very similar, of course we cannot exclude significant differences in the crystalline binding forces. In Fig. 2, the results of measurements of the heat capacity of single crystalline RhCl$_3$ are compared with the temperature dependence of $C/T$ of α-RuCl$_3$.

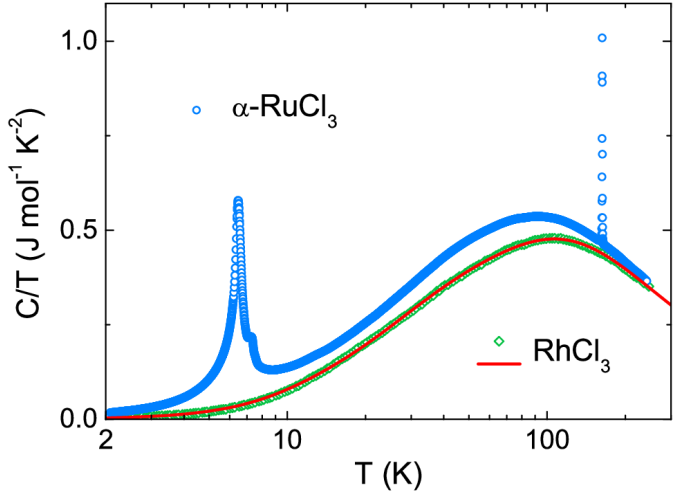

FIG. 2. Temperature dependence of the heat capacities of α-RuCl$_3$ (open blue circles) and of RhCl$_3$ (open green diamonds) plotted as $C/T$ vs. temperature up to 300 K on a semi-logarithmic plot. The measured heat capacity of the rhodium compound is fitted by a Debye-Einstein model (red solid line) as described in the text.

One naively would expect that the phonon-derived heat capacities of the rhodium and ruthenium compounds are similar, as the atomic masses are comparable. However, experimentally this is not observed. Despite the fact that the compounds plotted in Fig. 2 are isostructural and that mass renormalization should be marginal, we observe significant differences in the temperature dependence of the heat capacity up to 200 K, clearly signaling changes in bonding strength and concomitantly in the eigenfrequencies of phonon excitations. Of course, in the non-magnetic rhodium compound the lambda-like anomaly indicating the magnetic phase transition in α-RuCl$_3$ is missing, but the rhodium compound obviously also exhibits no structural phase transition as observed in a variety of tri-halides.

It seems interesting to note that the heat capacity measured for the rhodium compound, at low temperatures strictly follows a $T^3$ dependence, expected for a three-dimensional (3D) lattice, despite the fact that this is a strongly layered 2D compound with very weak inter-layer vdW interactions. The specific-heat data documented in Fig. 2 also show that the differences between ruthenium and the non-magnetic isostructural reference compound are too large to be attributed to the different masses of rhodium and ruthenium and the magnetic degrees of freedom alone.

To adequately tackle the problem of the correct phonon background we performed ab-initio phonon calculations of both isostructural compounds. These are detailed in the Appendix A: Figure 5 documents the wave-vector dependence of the calculated phonon eigenfrequencies along all high-symmetry directions and the resulting density of states of both compounds. We found that - despite the similar mass ratios - the eigenfrequencies of the rhodium compound are significantly enhanced. This results from the fact that due to enhanced chemical bonding the volume of the rhodium compound becomes reduced and the eigenfrequencies harden considerably (see Fig. 5 of Appendix A). From the phonon density of states (DOS) the temperature dependencies of the heat capacities for both compounds were calculated. The results are shown in Fig. 6 of the Appendix B. In the temperature range investigated, the heat capacity RhCl$_3$ is well below that of α-RuCl$_3$, indicative of a higher Debye temperature of the rhodium compound. Temperature scaling by a factor of 0.92 is necessary to map the heat capacity of the rhodium compound on the isostructural ruthenium compound in the investigated temperature range. The empty circles in Fig. 6 in the Appendix B document that this scaling works rather well and hence, that sizable renormalization of the phonon background is needed to describe the lattice derived heat capacity of α-RuCl$_3$.

In order to get an appropriate phonon background we fitted the heat capacity of the RhCl$_3$ by a Debye-Einstein model utilizing two Debye and two Einstein modes with a total of roughly 12 degrees of freedom (DoF). The solid line in Fig. 2 shows the result of this fit. The eigenfrequencies and DoF used for this model fit are documented in Table 1 of the Appendix B. This fit as documented in Fig. 2 was temperature-scaled by a factor of 0.92 and has been taken as appropriate phonon background for α-RuCl$_3$ for zero external magnetic field and for all fields up to 9 T.

The excess heat capacity of α-RuCl$_3$ as determined using this procedure is shown in Fig. 3 for three different external magnetic fields. Turning first to zero external field (black spheres), we find the anomaly due to the magnetic phase transition at ~ 6 K and a broad hump located around ~ 70 K. The λ-type anomaly certainly results from canonical spin degrees of freedom. The hump at 70 K could be due to the release of entropy due to itinerant Majorana fermions. The temperature scale is of the order of the expected Kitaev exchange in the title compound. This behavior resembles the proposed theoretical crossover scenario from a magnon-like behavior at low temperatures to fractionalized spin degrees of freedom at elevated temperatures [23].

For increasing in-plane magnetic fields, long-range spin order becomes continuously suppressed, with the heat capacity anomaly being reduced in height, becoming broader and shifted towards low temperatures, while the high-temperature hump remains essentially unchanged. Beyond quantum criticality a broad hump evolves at temperatures close to 10 K. Taking the excess heat capacity at 9 T as documented in Fig. 3 (green spheres), we find a two-peak structure of the heat capacity with the two anomalies at ~ 10 K and ~ 70 K. The temperature scale of the low-temperature peak is too high compared to model predictions. The high-temperature anomaly, which is expected at temperatures corresponding to the Kitaev exchange, is not too far from the theoretical expectations.

The entropy release at the low and high temperature peaks will be a further crucial test of the nature of these two anomalies. To determine the entropy, Fig. 7 in the Appendix B shows the excess heat capacity vs. temperature plotted as $\Delta C/T$ vs. $T$ for all magnetic fields investigated. The temperature evolution of the entropy at 0, 5.5 and 9 T is plotted in the inset of Fig. 3 and shows a clear two-step increase as theoretically predicted. Astonishingly, even in zero external field the entropy of the heat-capacity anomaly at the onset of magnetic order is well below $R\ln2$ expected for the ordering of a spin $S = ½$ system. The entropy due to magnetic order is very little affected by the external field and remains, even when long-range magnetic order is fully suppressed. The entropy release of the high-temperature peak is very close to $R/2 \ln2$ as predicted for a KSL and seems independent of magnetic field. The entropy released as latent heat, characteristic of the first-order structural phase transition is very small. However, it is interesting to note that even at the highest temperatures (> 200 K), the entropy release expected for a spin $S = ½$ system is not fully recovered (see inset of Fig. 3).

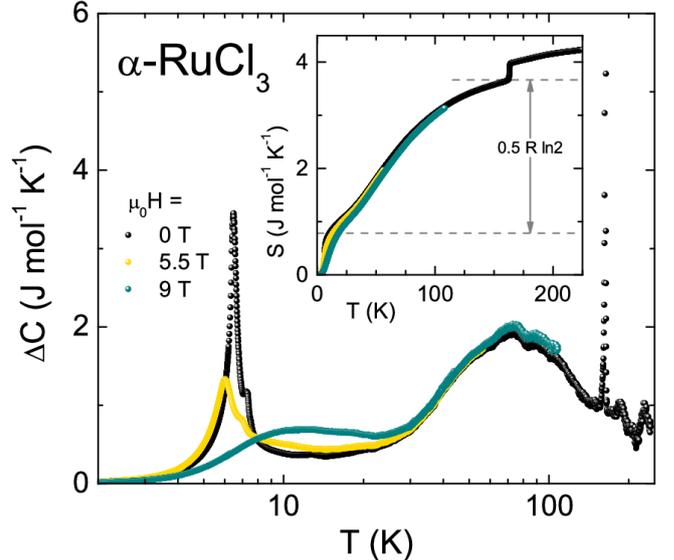

FIG. 3. Semi-logarithmic plot of the temperature dependence of the excess heat capacities of α-RuCl$_3$ plotted as $C/T$ vs. temperature for a series of in-plane magnetic fields as indicated in the figure. The excess heat capacity was obtained by subtracting phonon contributions as determined from the RhCl$_3$ measurements, which were temperature-scaled by a factor of 0.92 (for details see Appendix A). The inset shows the temperature dependence of the entropy up to 225 K. While the low-temperature anomaly carries only marginal weight, the high-temperature anomaly comes close to $R/2 \ln2$ expected for the entropy release due to itinerant Majorana fermions.

From heat capacity and entropy, the low-temperature peak seems to be connected with canonical spin-degrees of freedom close to magnetic order. The λ-like anomaly at small fields due to the onset lo long-range order, beyond quantum criticality transforms into a broad peak with roughly the same amount of entropy. As has been documented by THz spectroscopy by Wang et al. [18], magnetic excitations of similar energies are observed above and below quantum criticality. The nature of these excitations is a matter of dispute and deserves further investigations.

It seems reasonable to assume that the high-temperature peak results from itinerant Majorana fermions. The temperature scale and the entropy release are of the correct order and close to theoretical predictions. Broad continua have been observed in recent THz experiments centered around 6 meV, which survive up to 9 T [18] and are stable at least up to 150 K [37]. Both observations in good agreement with the thermodynamic results documented in Fig. 3. This figure provides some experimental evidence that at elevated temperatures, throughout the paramagnetic regime in a broad range of magnetic fields, the KSL is the stable quantum state in α-RuCl$_3$. While the low-temperature anomaly signals spin ordering or conventional fluctuating moments, we find a crossover to fractionalized excitations at higher temperatures.

Finally, we would like to have a closer look on the temperature dependence of α-RuCl$_3$ around the magnetic phase transition, also as function of external magnetic fields. These results are shown in Fig. 4(a) where we plotted the heat capacity between 2 and 10 K for in-plane magnetic fields between 0 and 9 T. At zero external field, the heat-capacity anomaly is located close to 6.4 K. There appears a small anomaly at 7.2 K. This result could indicate minor stacking disorder. However, so far all reports about magnetic ordering temperatures due to stacking faults indicate significantly higher ordering temperatures. Close to 7 T the magnetic phase transition becomes fully suppressed, leaving a broad hump shifted to slightly higher temperatures.

A closer inspection of the temperature dependence of the AFM anomaly shows that the magnetic phase transition at 1T exhibits a somewhat stronger shift towards lower temperatures and shifts to slightly higher temperatures again for further increasing magnetic fields. The field dependence of the magnetic phase transition for in-plane fields is indicated in the inset if Fig. 4(a). Here we plotted the field dependencies of both anomalies starting at 6.4 and 7.2 K in zero field. The fact that these anomalies, as function of temperature and magnetic field behave very similar, provide arguments that these anomalies are of similar nature and are related to domains with slightly different stacking order. For both anomalies, in the field dependence we observe a small local minimum close to external fields of ~ 1 – 2 T. It is unclear if these anomalies signal a critical or characteristic temperature of α-RuCl$_3$ in external fields. In literature, there are reports about a susceptibility anomaly for in-plane fields close to 1.2 T [43] and an anomaly in neutron scattering studies due to a redistribution of domains also in low fields [41]. Remarkably, close to 8 K in the temperature dependence of the excess heat capacity we observe a characteristic crossing point where all heat capacity curves meet as function of magnetic field. Similar crossing points have been observed in a variety of different materials and seem to be a characteristic signature of correlated systems [51].

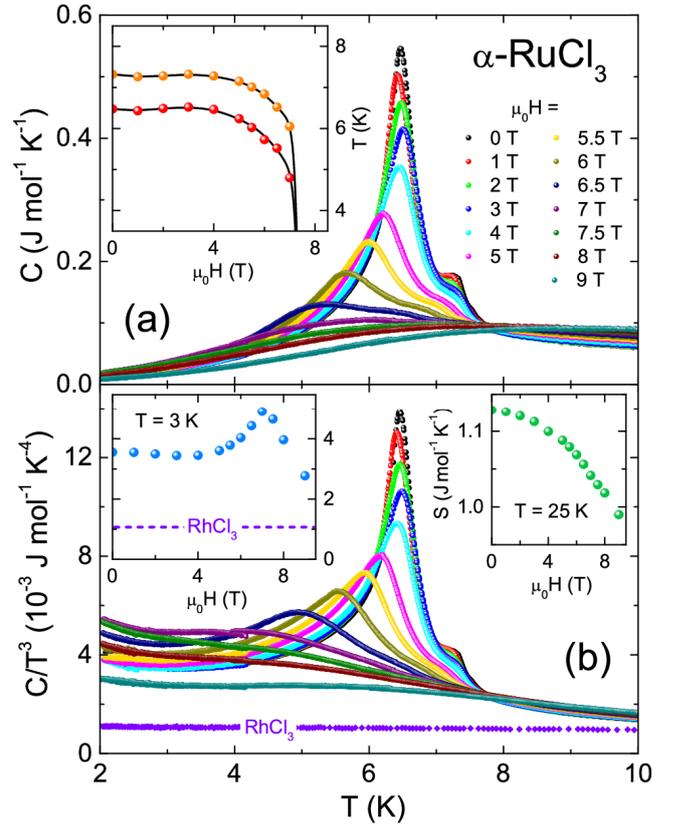

FIG. 4. Temperature dependence of the heat capacities of α-RuCl$_3$ at temperatures around the magnetic phase transition for a series of in-plane magnetic fields as indicated in the figure. (a) Temperature dependence of heat capacity as measured for external fields between 0 and 9 T. The inset shows the field dependence of the ordering temperature. (b) Temperature dependence of heat capacity for the same magnetic fields plotted as $C/T^3$. The heat capacity of RhCl$_3$ is also indicated. The left upper inset in (b) shows the field dependence of $C/T^3$ determined at 3 K. The right upper inset in (b) shows the field dependence of the entropy at 25 K.

Figure 4(b) shows the same data, plotted as $C/T^3$ vs. $T$. The results of α-RuCl$_3$ are compared with the single-crystalline data as measured in RhCl$_3$ at zero external field. Turning first to the rhodium results: In the temperature regime shown, the heat capacity is dominated by a pure $T^3$ Debye behavior of a 3D crystal, excluding any speculation about a possible 2D behavior of these vdW layered materials. This is also documented by the phonon fits documented in

Fig. 2. In the full temperature range investigated, the ruthenium heat capacity is always significantly enhanced compared to the phonon background pointing towards an enhanced Debye temperature of the rhodium compound. Even at the highest magnetic fields with fully suppressed magnetic order, the heat capacity of α-RuCl3 is still strongly enhanced signaling the importance of magnetic contributions beyond quantum criticality. In the AFM phase, this enhancement can be attributed to the contributions from magnetic excitations, which in a gapless AFM also will follow a $T^3$ dependence. With decreasing ordering temperatures, the magnetic contributions increase, become maximal at the critical field and decrease for further increasing fields [see left upper inset in Fig. 4(b)]. This observation provides further support for the interpretation that the 10 K anomaly for fields > 7 T continuously evolves from the λ-like anomaly at the onset of magnetic order in low fields. Astonishingly the temperature dependence of the specific heat at all magnetic fields is close to a $T^3$ law for all external magnetic in-plane fields investigated. This signals the importance of AFM excitations, with zero or a small energy gap for $T < T_N$ and for decreasing AFM-type fluctuations for fields beyond quantum criticality up to 9 T.

The scaling of the low-temperature heat capacity has been studied in detail by Wolter et al. [43] down to temperatures of 0.4 K. These authors found a temperature scaling with a power of 2.5 and they identified a strongly temperature-dependent energy gap, almost closing at the quantum-critical point. The heat capacity of α-RuCl3 in the sub-Kelvin regime has been reported by Kasahara et al. [44]. These authors found roughly a $T^3$ scaling of the heat capacity just beyond quantum criticality. It has to be clearly stated that neither our data nor the results by Wolter et al. [43] and Kasahara et al. [44] provide experimental evidence for the existence of a peak in the specific heat at low temperatures, close to 2 K as theoretically predicted [5,9,10]. At this temperature, one expects that excitations of localized Majorana fermions release an entropy of order $R/2 \ln2$. This amount of entropy is not hidden in the low-temperature excess heat capacity. The entropy up to the onset of magnetic order at 6 K is always less than 0.2 J/(mol K) and hence a small fraction (< 10 %) of the expected entropy release due to localized Majorana fermions. This peak due to localized fluxes must be located at significantly lower temperatures, or has to be identified with the cusp at 10 K as shown in Fig. 3 for in-plane fields > 7 T.

## IV. SUMMARY AND CONCLUDING REMARKS

In conclusion, in this manuscript we provide detailed experimental results on the heat capacity of α-RuCl3 as function of temperature and external magnetic in-plane fields. The main aim of this work, to provide thermodynamic evidence of fractionalization of spin-degrees of freedom, crucially depends on the estimate of the intrinsic phonon background. To solve this problem, we measured the heat capacity of isostructural and non-magnetic RhCl3. The correct temperature-scaling factor has been determined utilizing a critical comparison of the heat capacities of the rhodium and the ruthenium compounds, which were determined by ab-initio calculations of the phonon properties and concomitant phonon densities of states. We show that the correct phonon background cannot be obtained by simple rescaling using the different atomic masses.

The main result of this work as presented in Fig. 3 can be summarized as follows: At zero external magnetic field, a λ-like anomaly at ~ 6 K, corresponding to the onset of long-range spin order, is followed by a high-temperature cusp at ~ 70 K, which we interpret as fingerprint of itinerant Majorana fermions. This zero-field result can directly be compared with the heat-capacity results documented in Ref. [13]. Despite a close over-all similarity of the results, differences concern the temperature range of the high-temperature anomaly (~ 100 K) and that these authors determined a roughly linear increase of the excess heat capacity between 50 K and 100 K.

At 9 T, beyond quantum criticality, we observe two cusps in the temperature dependence of the heat capacity of α-RuCl3, close to 10 K and close to 70 K. As function of increasing magnetic fields, the low-temperature anomaly continuously evolves from the specific-heat anomaly indicative of long-range magnetic order. The high-temperature anomaly remains almost constant as function of external fields and - from our point of view - signals excitations of itinerant fermions in the paramagnetic phase. This interpretation is supported by entropy considerations: The entropy released by the low-temperature anomaly is well below 1 J/(mol K). It is significantly too low for the entropy due to the ordering of a spin $S = \frac{1}{2}$ system, but it is also too low when compared to the predicted entropy release of localized fluxes in the framework of a KSL. The high-temperature anomaly signals an entropy release close to $R/2 \ln 2$, expected for itinerant Majorana fermions.

Summarizing the obtained results, we conclude that at zero external magnetic field, magnetic excess heat capacity at the onset of long-range magnetic order is followed by contributions of fractionalized excitations at higher temperatures. Beyond quantum criticality, the low-temperature anomaly still resembles contributions from canonical spin-degrees of freedom, while the high-temperature anomaly remains unchanged. A natural explanation for these observations has been presented in a recent work by Rousochatzakis et al. [23]: These authors have shown that, while magnon-like response - characteristic for conventional magnetic order - appears at low temperatures, the characteristics of a KSL evolves at elevated temperatures. It seems that this crossover behavior is documented by our specific heat experiments in zero magnetic fields as well as beyond quantum criticality.

## ACKNOWLEDGEMENTS

This work has been partly supported by the Deutsche Forschungsgemeinschaft (DFG) via the Transregional Collaborative Research Center TRR 80. The work of D. A. P. and V. G. M. was supported by Act 211 Government of the Russian Federation, contract Nr. 02.A03.21.0006. A. A. T. was supported by the Federal Ministry of Education and Research through the Sofja Kovalevskaja Award of the Alexander von Humboldt Foundation.

S.W. performed the heat-capacity experiments, V. T. synthesized the samples, D. A. P, V. G. M. and A. T. performed the ab-initio phonon calculations. A. L. wrote the manuscript with contributions from all authors.

## APPENDIX A: AB-INITIO PHONON CALCULATIONS

Phonon spectra and lattice contributions to the specific heat of α-RuCl$_3$ and RhCl$_3$ were calculated using the frozen-phonon method implemented in the PHONOPY code [50]. Atomic displacements of 0.01 Å were used to induce non-zero forces in the 2 x 1 x 2 supercell containing 64 atoms. The forces were calculated in the VASP code within the projector-augmented wave (PAW) method [52,53] using the Perdew-Burke-Ernzerhof (PBE) flavor of the exchange-correlation potential [54].

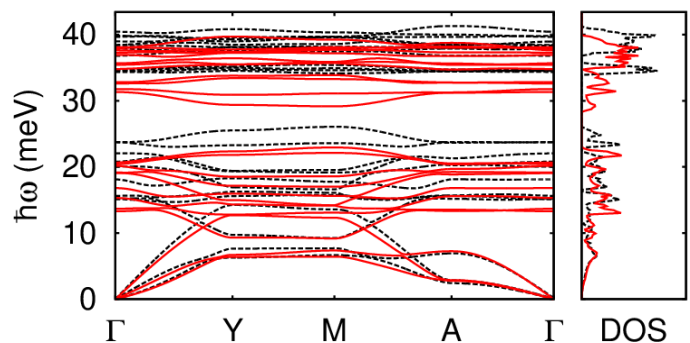

Fig.5. Phonon eigenfrequencies of α-RuCl$_3$ (red solid lines) and of RhCl$_3$ (black dashed lines) vs. eigenvectors along all high-symmetry directions in reciprocal space and the energy dependence of the phonon density of states (DOS) as derived from this phonon dispersion.

Initial calculations revealed imaginary phonon frequencies and a tendency toward dimerization in α-RuCl$_3$ when calculations were performed on the PBE level. Therefore, electronic correlations in the Ru (Rh) 4d shell and the spin-orbit coupling were additionally taken into account. Electronic correlations were treated within the DFT+U scheme, using the on-site Coulomb repulsion $U_d = 1.8$ eV and the Hund's coupling $J_d = 0.4$ eV [55,56].

The resulting phonon eigenfrequencies as function of wave vectors along all main symmetry directions of the reciprocal lattice are shown for both compounds in Fig. 5. The phonon modes of these layered vdW compounds are similar and are grouped in well separated frequency regimes. In addition to acoustic modes, two groups of optical modes can be identified. The lowest acoustic modes appear close to the A point of the Brillouin zone and are close to 2 and 6 meV. All phonon modes are well below 45 meV. For the majority of phonon branches, the eigenfrequencies of the phonon modes of RhCl$_3$ are significantly enhanced compared to the ruthenium compound. This behavior is mirrored also by the phonon density of states documented in Fig. 5.

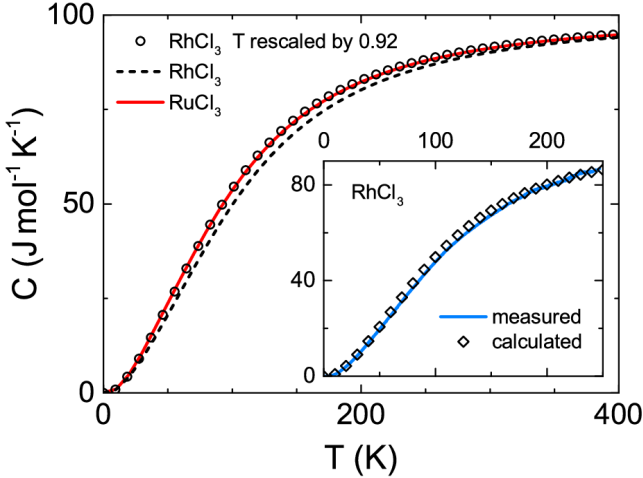

FIG. 6. Calculated temperature dependencies of the heat capacities of α-RuCl$_3$ (red line) and RhCl$_3$ (black dashed line). The two heat capacities can be matched by temperature scaling of the RhCl$_3$ heat capacity by a factor of 0.92 (empty circles). The inset documents that the phonon-derived heat capacity of RhCl$_3$ as determined from ab-initio methods (open diamonds) is very close to the experimental results (blue solid line).

From the phonon densities of states the temperature dependence of the heat capacities for both compounds were calculated. The results are shown in Fig. 6. α-RuCl$_3$ exhibits significantly smaller phonon frequencies compared to RhCl$_3$, resulting in a reduced mean Debye temperature and therefore a significantly enhanced heat capacity up to room temperature. Close to 400 K, the heat capacities of both compounds approach the high-temperature Dulong-Petit limit. By temperature scaling the heat capacity of RhCl$_3$ by a factor of 0.92 (empty circles in Fig. 6) a perfect match to the heat capacity of α-RuCl$_3$ is achieved. This temperature-scaling factor, which was determined by ab-initio phonon calculations, was used to estimate the intrinsic phonon contribution of the ruthenium compound.

## APPENDIX B: HEAT CAPACITY AND ENTROPY

The heat capacity of the rhodium compound as shown in Fig. 2 (and in the inset of Fig. 6) was fitted utilizing a Debye-Einstein model using two Debye and two Einstein modes. The resulting fit parameters including phonon eigenfrequencies and degrees of freedom (DoF) of these modes are indicated in Table 1. The result of this fit is documented in Fig. 2 as red solid line. All the phonon frequencies are below 45 meV as determined in the ab-initio frozen-phonon method. The total DoF comes close to a value of 12 expected for the 4 atoms per unit cell of the title compound.

Table 1: Fit parameters of the Debye-Einstein model used to calculate the heat capacities of RhCl$_3$. For the fits documented in Fig. 2 of the manuscript, two Debye (D) and two Einstein terms (E) have been used. The phonon eigenfrequencies ν are given in units of K. The degrees of freedom (DoF) for the different terms are indicated for each eigenfrequency.

|     | RhCl$_3$ | |
| --- | --- | --- |
|     | hν/k$_B$ (K) | DoF |
| D1  | 75  | 0.366 |
| D2  | 185 | 2.92 |
| E1  | 275 | 3.15 |
| E2  | 475 | 5.81 |

Figure 7 shows the excess heat capacity after subtracting the temperature-scaled phonon background determined from this fit to the rhodium measurements. The results are plotted as $\Delta C/T$ vs. temperature and are shown for all measured external magnetic fields between 0 and 9 T. The λ-like anomaly at the onset of magnetic order becomes continuously suppressed and remains as hump-like anomaly with a maximum close to 5 K for external fields approaching the quantum-critical transition. This is documented in Fig. 4 of the main text. Beyond the quantum-critical point, the cusp-like phenomenon abruptly is shifted to significantly higher temperatures, despite the fact that the peak signaling the onset of long-range magnetic order continuously becomes suppressed (see Fig. 4). Despite this fact, we observe a continuous evolution of the magnetic anomaly from a λ-peak at low fields, signaling the onset of long-range order into a hump-like shape at fields beyond the critical field signaling fluctuating moments. Hence, the interpretation of this low-temperature anomaly in terms of canonical spin degrees of freedom is most plausible.

The results as measured in external fields of 9 T (green spheres) nicely exhibit the characteristic two-peak structure with a relatively narrow low-temperature and a broad high-temperature peak. We would like to recall that in this field range the anomaly due to the onset of long-range or short-range magnetic order was shifted to low temperatures (see Fig. 4). The cusp-like anomaly, which appears beyond quantum criticality (> 7 T) is a continuation of the λ-like anomaly at low fields.

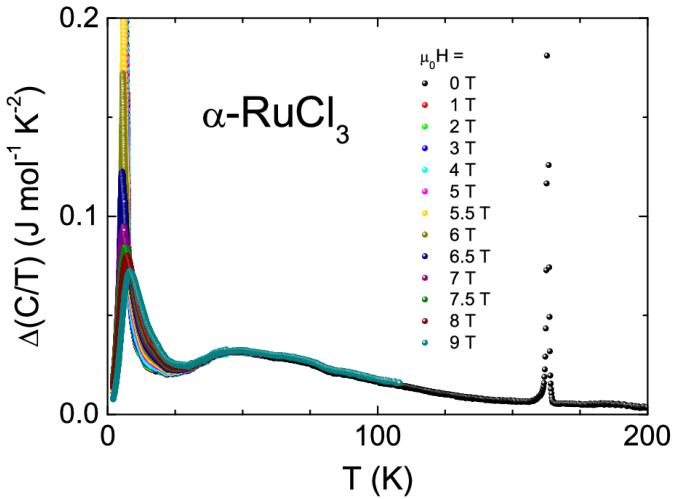

FIG. 7. Excess heat capacity of α-RuCl$_3$ plotted as $C/T$ vs. temperature for a series of magnetic in-plane fields between 0 and 9 T. Even beyond quantum criticality (> 7 T), a two-peak structure is clearly visible with peaks close to 10 and 40 K. For clarity the vertical scale has been cut at 0.2 J/(mole K$^2$). Despite the fact that the peak at 10 K looks like a reminder of the AFM phase transition, it obviously has changed character as long-range magnetic order is suppressed close to 7 T.

Using these results, the temperature release of the entropy has been calculated. The entropy release up to the highest measured temperatures is plotted in Fig. 3 of the main paper. For a clearer inspection of the low-temperature entropy release, the temperature dependence of the entropy up to 40 K is documented in Fig. 8. At zero external magnetic field the entropy release at the AFM phase transition is of order 0.7 J/(mol K) corresponding to ~ 15% of the entropy expected in the ordering process of a spin ½ system. On increasing magnetic fields this entropy value remains rather constant, only being slightly smeared out. The magnetic field dependence of the entropy release at 25 K is shown at the inset of Fig. 8. The entropy continuously decreases up to 9 T, however, only by a value of 10%. From this observation we conclude that the low-temperature anomaly in α-RuCl$_3$ is derived from canonical spin degrees of freedom, establishing long-range magnetic order for fields < 7 T and undergoing strong fluctuations in a magnetic system with short-range order beyond quantum criticality.

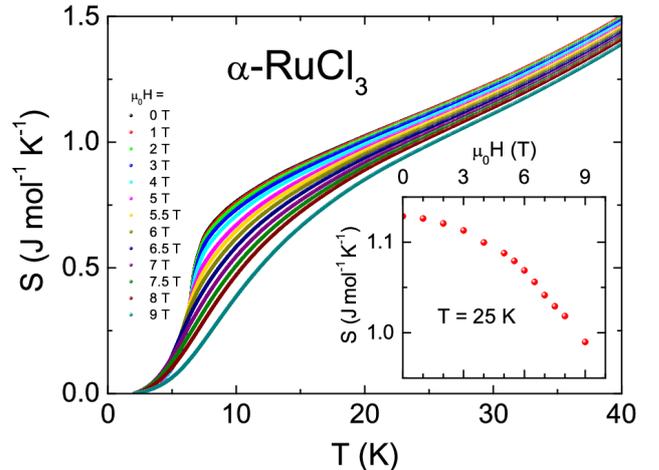

FIG. 8. Entropy vs. temperature in α-RuCl$_3$ at a series of magnetic in-plane fields between 0 and 9 T. The step-like increase at the onset of magnetic order becomes smeared out for external field beyond the critical field. The inset shows the entropy as function of magnetic field at 25 K.

## V. REFERENCES

[1] A. Kitaev, Anyons in an exactly solved model and beyond, Ann. Phys. **321**, 2 (2006).
[2] S. M. Winter, A. A. Tsirlin, M. Daghofer, J. van den Brinck, Y. Singh, P. Gegenwart, and R. Valenti, Models and materials for generalized Kitaev magnetism, J. Phys. Cond. Matter. **29**, 493002 (2017).
[3] M. Hermanns, I. Kimchi, and J. Knolle, Physics of the Kiaev model: fractionalization, dynamical correlations, and material connections, Ann. Rev. Cond. Matter Phys. **9**, 17 (2018).
[4] L. Balents, Spin liquids in frustrated magnets, Nature (London) **464**, 199 (2010).

[5] J. Nasu, M. Udagawa, and Y. Motome, Vaporization of Kitaev Spin Liquids, Phys. Rev. Lett. **113**, 197205 (2014).

[6] J. Nasu, M. Udagawa, and Y. Motome, Thermal fractionalization of quantum spins in a Kitaev model: Temperature-linear specific heat and coherent transport of Majorana fermions, Phys. Rev. B **92**, 115122 (2015).

[7] J. Yoshitake, J. Nasu, and Y. Motome, Fractional Spin Fluctuations as a Precursor of Quantum Spin Liquids: Majorana Dynamical Mean-Field Study for the Kitaev Model, Phys. Rev. Lett. **117**, 157203 (2016).

[8] Y. Yamaji, T. Suzuki, T. Yamada, S.-I. Suga, N. Kawashima, and M. Imada, Clues and criteria for designing a Kitaev spin liquid revealed by thermal and spin excitations of the honeycomb iridate $Na_2IrO_3$, Phys. Rev. B **93**, 174425 (2016).

[9] A. Catuneanu, Y. Yamaji, G. Wachtel, Y. B. Kim, and H.-Y. Kee, Path to stable quantum spin liquids in spin-orbit coupled correlated materials, npj Quantum Materials **3**, 23 (2018).

[10] A. M. Samarakoon, G. Wachtel, Y. Yamaji, D. A. Tennant, C. D. Batista, and Y.-B. Kim, Classical and quantum spin dynamics of the honeycomb model, Phys. Rev B **98**, 045121 (2018).

[11] A. Banerjee, C. A. Bridges, J.-Q. Yan, A. A. Aczel, L. Li, M. B. Stone, G. E. Granroth, M. D. Lumsden, Y. Yiu, J. Knolle, S. Bhattacharjee, D. L. Kovrizhin, R. Moessner, D. A. Tennant, D. G. Mandrus, and S. E. Nagler, Proximate Kitaev quantum spin liquid behaviour in a honeycomb magnet, Nat. Mater. **15,** 733 (2016).

[12] K. Ran, J. Wang, W. Wang, Z.-Y. Dong, X. Ren, S. Bao, S. Li, Z. Ma, Y. Gan, Y. Zhang, J. T. Park, G. Deng, S. Danilkin, S.-L. Yu, J.-X. Li, and J. Wen, Spin-Wave Excitations Evidencing the Kitaev Interaction in Single Crystalline α-RuCl₃, Phys. Rev. Lett. **118,** 107203 (2017).

[13] S.-H. Do, S.-Y. Park, J. Yoshitake, J. Nasu, Y. Motome, Y. S. Kwon, D. T. Adroja, D. J. Voneshen, K. Kim, T.-H. Jang, J.-H. Park, K.-Y. Choi, and S. Ji, Majorana fermions in the Kitaev quantum spin system α-RuCl₃**,** Nat. Phys. **13,** 1079 (2017).

[14] A. Banerjee, P. Lampen-Kelley, J. Knolle, C. Balz, A. A. Aczel, B. Winn, Y. Liu, P. Pajerowski, J. Yan, C. A. Bridges, A. T. Savici, B. C. Chakoumakos, M. D. Lumsden, D. A. Tennant, R. Moessner, D. G. Mandrus, and S. E. Nagler, Excitations in the field-induced quantum spin liquid state of α-RuCl₃, npj Quantum Mater. **3,** 8 (2018).

[15] L. J. Sandilands, Y. Tian, K. W. Plumb, Y.-J. Kim, and K. S. Burch, Scattering Continuum and Possible Fractionalized Excitations in α-RuCl₃, Phys. Rev. Lett. **114,** 147201 (2015).

[16] J. Nasu, J. Knolle, D. L. Kovrizhin, Y. Motome, and R. Moessner, Fermionic response from fractionalization in an insulating two-dimensional magnet, Nat. Phys. **12,** 912 (2016).

[17] A. Glamazda, P. Lemmens, S.-H. Do, Y. S. Kwon, and K.-Y. Choi, Relation between Kitaev magnetism and structure in α-RuCl₃, Phys. Rev. B **95**, 174429 (2017).

[18] Zhe Wang, S. Reschke, D. Hüvonen, S.-H. Do, K.-Y. Choi, M. Gensch, U. Nagel, T. Room, and A. Loidl, Magnetic Excitations and Continuum of a Possibly Field-Induced Quantum Spin Liquid in α-RuCl₃, Phys. Rev. Lett. **119**, 227202 (2017).

[19] A. Little, L. Wu, P. Lampen-Kelly, A. Banerjee, S. Patankar, D. Rees, C. A. Bridges, J.-Q. Yan, D. Mandrus, S. E. nagler, and J. Orenstein, Antiferromagnetic resonance and terahertz continuum in α-RuCl₃, Phys. Rev. Lett. **119**, 227201 (2017).

[20] L. Janssen, E. C. Andrade, and M. Vojta, Magnetization processes of zigzag states on the honeycomb lattice: Identifying spin models for α-RuCl₃ and $Na_2IrO_3$, Phys. Rev. B **96**, 064430 (2017).

[21] T. Suzuki and S. Suga, Effective model with strong Kitaev interactions for α-RuCl₃, Phys. Rev. B **97**, 134424 (2018).

[22] A. Pidatella, A. Metavitsiadis, and W. Brenig, Heat transport in the anisotropic Kitaev spin liquid, unpublished, arXiv: 1810.04674

[23] I. Rousochatzakis, S. Kourtis, J. Knolle, R. Moessner, and N. B. Perkins, Quantum spin liquid at finite temperature: proximate dynamics and persistent typicality, unpublished, arXiv: 1811.01671

[24] S. M. Winter, K. Riedl, P. A. Maksimov, A. L. Chernyshev, A. Honecker, and R. Valenti, Breakdown of magnons in a strongly spin-orbital coupled magnet, Nature Commun. **8**, 1152 (2017).

[25] S. M. Winter, K. Riedl, D. Kaib, R. Coldea, and R. Valenti, Probing α-RuCl₃ beyond magnetic order: Effects of temperature and magnetic field, Phys. Rev. Lett. **120**, 077203 (2018).

[26] J. A. Sears, M. Songvilay, K. W. Plumb, J. P. Clancy, Y. Qiu, Y. Zhao, D. Parshall, and Y.-J. Kim, Magnetic order in α-RuCl3: A honeycomb-lattice quantum magnet with strong spin-orbit coupling, Phys. Rev. B **91**, 144420 (2015).

[27] M. Majumder, M. Schmidt, H. Rosner, A. A. Tsirlin, H. Yasuoka, and M. Baenitz, Anisotropic $Ru^{3+}$ $4d5$ magnetism in the α-RuCl3 honeycomb system: Susceptibility, specific heat, and zero-field NMR, Phys. Rev. B **91**, 180401(R) (2015).

[28] H. B. Cao, A. Banerjee, J.-Q. Yan, C. A. Bridges, M. D. Lumsden, D. G. Mandrus, D. A. Tennant, B. C. Chakoumakos, and S. Nagler, Low-temperature crystal and magnetic structure of α-RuCl₃, Phys. Rev. B **93**, 134423 (2016).

[29] S.-Y. Park, S.-H. Do, K.-Y. Choi, D. Jang, T.-H. Jang, J. Schefer, C.-M. Wu, J. S. Gardner, M. S. Park, J.-H. Park, and S. Ji, Emergence of the Isotropic Kitaev Honeycomb Lattice with Two-dimensional Ising Universality in α-RuCl₃, unpublished, arXiv: 1609.05690 (2016).

[30] Y. Kubota, Hi. Tanaka, T. Ono, Y. Narumi, and K. Kindo, Successive magnetic phase transitions in α-RuCl3: XY-like frustrated magnet on the honeycomb lattice, Phys Rev. B **91**, 094422(R) (2015).

[31] D. Hirobe, M. Sato, Y. Shiomi, H. Tanaka, and Eiji Saitoh, Magnetic thermal conductivity far above the Néel temperature in the Kitaev-magnet candidate α-RuCl₃, Phys. Rev B **95**, 241112(R) (2017).

[32] K. Brodersen, F. Moers, and H. G. Schnering, Zur Struktur des Iridium(III) und Ruthenium(III)-chlorids, Naturwissenschaften **52**, 205 (1965).

[33] K. Brodersen, G. Thiele, H. Ohnsorge, I. Recke, and F. Moers, Die Struktur des $IBr_3$ und über die Ursachen der Fehlordnungserscheinungen bei den Schichtstrukturen der kristallisierenden Edelmetallhalogeniden, J. Less-Common Met. **15**, 347 (1968).

[34] H.-J. Cantow, H. Hillebrecht, S. N. Magonov, H. W. Rotter, M. Drechsler, and G. Thiele, Atomic Structure and Superstructure of α-RuCl₃ by Scanning Tunneling Microscopy**,** Angew. Chem. Int. Ed. **29**, 537 (1990).

[35] S. Reschke, F. Mayr, Zhe Wang, Seung-Hwan Do, K.-Y. Choi, and A. Loidl, Electronic and phonon excitations in α-RuCl3, Phys. Rev. B **96**, 165120 (2017).

[36] M. He, X. Wang, L. Wang, F. Hardy, T. Wolf, P. Adelmann, T. Brückel, Y. Su, and Ch. Meingast, Uniaxial and hydrostatic pressure effects in α-RuCl3 single crystals via thermal-expansion measurements, unpublished, arXiv: 1712.08511 (2017).

[37] S. Reschke, F. Mayr, S. Widmann, H.-A. Krug von Nidda, V. Tsurkan, M. V. Eremin, S.-H. Do, K.-Y. Choi, Zhe Wang, and A. Loidl, Sub-gap optical response in the Kitaev spin-liquid candidate α-RuCl₃, J. Phys.: Condens. Matter **30**, 475604 (2018).

[38] R. D. Johnson, S. C. Williams, A. A. Haghighirad, J. Singleton, V. Zapf, P. Manuel, I. I. Mazin, Y. Li, H. O. Jeschke, R. Valentí, and R. Coldea, Monoclinic crystal structure of α-RuCl3 and the zigzag antiferromagnetic ground state, Phys. Rev. B **92,** 235119 (2015).

[39] M. Ziatdinov, A. Banerjee, A. Maksov, T. Berlijn, W. Zhou, H. B. Cao, J.-Q. Yan, C. A. Bridges, D. G. Mandrus, S. E. Nagler, A. P. Baddorf, and S. V. Kalinin, Atomic-scale observation of structural and electronic orders in the layered compound α-RuCl$_3$, Nat. Commun. **7**, 13774 (2016).

[40] S.-H. Baeck, S.-H. Do, K.-Y. Choi, Y. S. Kwon, A. U. B. Wolter, S. Nishimoto, J. van den Brink, and B. Büchner, Evidence for a Field-Induced Quantum Spin Liquid in α-RuCl$_3$, Phys. Rev. Lett. **119**, 037201 (2017).

[41] J. A. Sears, Y. Zhao, Z. Xu, J. W. Lynn, and Y.-J. Kim, Phase diagram of α-RuCl$_3$ in an in-plane magnetic field, Phys. Rev. B **95**, 180411(R) (2017).

[42] J. Zheng, K. Ran, T. Li, J. Wang, P. Wang, B. Liu, Z. Liu, B. Normand, J. Wen, and W. Yu, Gapless Spin Excitations in the Field-Induced Quantum Spin Liquid Phase of α-RuCl$_3$, Phys. Rev. Lett. **119**, 227208 (2017).

[43] A. U. B. Wolter, L. T. Corredor, L. Jannsen, K. Nenkov, S. Schönecker, S.-H. Do, K.-Y. Choi, R. Albrecht, J. Hunger, T. Coert, M. Vojta, and B. Büchner, Field-induced quantum criticality in the Kitaev system α-RuCl$_3$, Phys. Rev. B **96**, 041405(R) (2017).

[44] Y. Kasahara, T. Ohnishi, Y. Mizukami, O. Tanaka, S. Ma, K. Sugii, N. Kurita, H. Tanaka, J. Nasu, Y. Motome, T. Shibauchi, and Y. Matsuda, Majorana quantization and half-integer thermal quantum Hall effect in a Kitaev spin liquid, Nature **559**, 227 (2018).

[45] H. Bärnighausen and B. K. Handa, Die Kristallstruktur von Rhodium(III)-chlorid, J. Less Common Metals **6**, 226 (1964).

[46] B. Morosin and A. Narath, X-Ray Diffraction and Nuclear Quadrupole Resonance Studies of Chromium Trichloride, J. Chem. Phys. **40,** 1958 (1964).

[47] M. A. McGuire, H. Dixit, V. R. Cooper, and B. C. Sales, Coupling of crystal structure and magnetism in the layered, ferromagnetic insulator CrI$_3$, Chem. Mater. **27,** 612 (2015).

[48] M. A. McGuire, G. Clark, Santosh KC, W. M. Chance, G. E. Jellison Jr., V. R. Cooper, X. Xu, and B. C. Sales, Magnetic Behavior and Spin-Lattice Coupling in Cleavable, van der Waals Layered CrCl$_3$ Crystals, unpublished, arXiv:1706.01796 (2017).

[49] M. A. McGuire, J. Yan, P. Lampen-Kelley, A. F. May, V. R. Cooper, L. Lindsay, A. Puretzky, L. Liang, Santosh KC, E. Cakmak, S. Calder, and B. C. Sales, High temperature magnetostructural transition in van der Waals-layered α-MoCl$_3$, unpublished, arXiv: 1711.02708 (2017).

[50] A. Togo and I. Tanaka, First principles phonon calculations in materials science, Sci. Mater. **108**, 1 (2015).

[51] D. Vollhardt, Characteristic Crossing Points in Specific Heat Curves of Correlated Systems, Phys. Rev. Lett. **78**, 1307 (1997).

[52] G. Kresse and J. Furthmüller, Efficiency of ab-initio total energy calculations for metals and semiconductors using a plane-wave basis set, Comput. Mater. Sci. **6**, 15 (1996).

[53] G. Kresse and J. Furthmüller, Efficient iterative schemes for ab initio total-energy calculations using a plane-wave basis set, Phys. Rev. B **54**, 11169 (1996).

[54] J. P. Perdew, K. Burke, and M. Ernzerhof, Generalized gradient approximation made simple, Phys. Rev. Lett. **77**, 3865 (1996).

[55] H.-S. Kim, V. Shankar, A. Catuneanu, and H.-Y. Kee, Kitaev magnetism in honeycomb RuCl$_3$ with intermediate spin-orbit coupling, Phys. Rev. B **91**, 241110(R) (2015).

[56] T. Biesner, S. Biswas, W. Li, Y. Saito, A. Pustogow, M. Altmeyer, A.U.B. Wolter, B. Büchner, M. Roslova, T. Doert, S.M. Winter, R. Valenti, and M. Dressel, Detuning the honeycomb of α-RuCl$_3$: Pressure-dependent optical studies reveal broken symmetry, Phys. Rev. B **97**, 220401(R) (2018).